# HABITABLE ZONE PREDICTIONS AND HOW TO TEST THEM

*A white paper submitted to the Astro 2020 Decadal Survey on Astronomy and Astrophysics*

**Lead author:**
Ramses M. Ramirez
Earth-Life Science Institute (ELSI), Tokyo Institute of Technology, Tokyo 152-8550
Space Science Institute
email: rramirez@elsi.jp, rramirez@spacescience.org
phone: 03-5734-2183

**Co-authors:** Dorian S. Abbot (University of Chicago), Yuka Fujii (ELSI), Keiko Hamano (ELSI), Edwin Kite (University of Chicago), Amit Levi (Harvard University), Manasvi Lingam (Harvard University), Theresa Lueftinger (University of Vienna)**,** Tyler D. Robinson (Northern Arizona University), Andrew Rushby (University of California, Irvine), Laura Schaefer (Stanford University), Elizabeth Tasker (JAXA), Giovanni Vladilo (INAF-Osservatorio Astronomico di Trieste), Robin D. Wordsworth (Harvard University)

**Co-signers:** Chuanfei Dong (Princeton University), Nicolas Iro (University of Vienna), Stephen Kane (University of California, Riverside), Joshua Krissansen-Totton (University of Washington), Matthieu Laneuville (ELSI), Mercedes López-Morales (Smithsonian Astrophysical Observatory), Rodrigo Luger (Flatiron Institute), Bertrand Mennesson (Jet Propulsion Laboratory), Joshua Pepper (Lehigh University), Edward Schwieterman (University of California, Riverside)

**Abstract:** The habitable zone (HZ) is the region around a star(s) where standing bodies of water *could* exist on the surface of a rocky planet. The classical HZ definition makes a number of assumptions common to the Earth, including assuming that the most important greenhouse gases for habitable planets are $CO_2$ and $H_2O$, habitable planets orbit main-sequence stars, and that the carbonate-silicate cycle is a universal process on potentially habitable planets. Here, we discuss these and other predictions for the habitable zone and the observations that are needed to test them. We also, for the first time, argue why A-stars may be interesting HZ prospects. Instead of relying on unverified extrapolations from our Earth, we argue that future habitability studies require first principles approaches where temporal, spatial, physical, chemical, and biological systems are dynamically coupled. We also suggest that next-generation missions are only the beginning of a much more data-filled era in the not-too-distant future, when possibly *hundreds - thousands* of HZ planets will yield the statistical data we need to go beyond just finding habitable zone planets to actually determining which ones are most likely to exhibit life.

**INTRODUCTION**

As the next generation of missions are being assessed for NASA's next decadal survey, scientists will need to ensure that the tools available to find potentially habitable planets - and perhaps signs of life - are capable for the task. The best navigational tool for finding potentially habitable planets remains the habitable zone (HZ) (e.g.,[1,2]), which is the region around a star(s) where standing bodies of liquid water could be stable on a planetary surface [3]. Space-based observatories employ the HZ to select potentially habitable planets for follow-up habitability assessment. The HZ could become increasingly useful for finding potentially habitable planets and learning about their atmospheres.

The classical HZ definition[2,4] makes a number of assumptions common to Earth that should be tested by upcoming missions. First, it posits that the key greenhouse gases on HZ planets are $CO_2$ and $H_2O$[2]. It also assumes that the carbonate-silicate cycle, which is thought to maintain habitability on our own planet over million-year timescales[5], is (or equivalent carbon-cycling mechanism on planets containing land) a universal process on other habitable planets. Other central assumptions are that HZ planets orbit main-sequence stars and that the temporal evolution of a star is not critical in the assessment of a planet's habitability (but also see ref:[6]). Although the notion that liquid water is critical to life is itself an assumption, the suitability of other solvents for life (e.g. methane, ammonia) remains highly-debated (e.g.[7,8]). Plus, planets with seas consisting of these other solvents are located far enough away from their stars that detecting life's signatures may be impossible with current and upcoming technology. Thus, the focus on liquid water seems appropriate for now [3].

However, several investigators have recently argued that the classical definition may be overly-simplistic and too Earth-centric to be applied to other planets (e.g. [3,9,10]). Indeed, several attempts have been made to redefine the HZ. These include (but are not limited to) assuming different greenhouse gas combinations (e.g. [11–13]), considering the pre- and post-main-sequence temporal evolution (e.g.[14–17]), and the potential habitability of ocean worlds in the absence of plate tectonics (e.g. [18–21]). Unfortunately, without observations, the utility of the classical HZ or of any alternative HZ definitions is unknown. Only observations, followed by improved theories, can shed light on these questions. Here, we discuss several predictions expected by both classical and alternative HZ definitions and outline what is needed to test them.

**HABITABLE ZONE TESTS AND PREDICTIONS**

*Is the carbonate-silicate cycle really universal?*

As predicted from the carbonate-silicate cycle, planets near the inner edge of the HZ have atmospheres dominated by $H_2O$-vapor whereas those towards the outer edge are $CO_2$-dominated[2]. This is because stellar fluxes increase toward the inner edge, which promotes more rainfall (plus the higher temperatures increase weathering rates) and $CO_2$ drawdown. In contrast, reduced stellar fluxes near the outer edge lead to less rainfall (and decreased weathering rates), which promotes a net accumulation of atmospheric $CO_2$ (e.g.,[5]).

However, this cycle has never been directly observed on our Earth and its applicability on the cosmic scale is unknown. For instance, if multi-bar $CO_2$ atmospheres characterized planets near the outer edge, polar $CO_2$ condensation would lead to atmospheric collapse[22]. Moreover, global supply-limited weathering can inhibit the carbonate-silicate cycle and disrupt habitability [23]. These studies then suggest that the classical HZ treatment of the carbonate-silicate cycle may be overly-optimistic and underscores the dangers of making deductions from our planet.

Testing the universality of the carbonate-silicate cycle requires that missions measure the atmospheric $CO_2$ pressures of HZ planets as a function of semi-major axis for different star types (e.g., [24]). Trends can be determined with a sufficiently large statistical sample. Thus, if after observing many such systems, no evidence of such a trend is found, the idea can be falsified. Otherwise, this would be the best evidence for a universal carbonate-silicate cycle.

Another related idea is whether ocean worlds, with interiors that are rich enough in water (tens of percent of their mass) to form high-pressure ices on the sea floor, are potentially habitable [19–21,25]. Lacking continents, the carbonate-silicate cycle would not operate on these worlds [26]. Volcanism would also be subdued by the higher interior pressures (e.g. [27]).

However, the carbonate-silicate cycle could be replaced by one in which $CO_2$ outgassing and ocean fluxes are balanced. Although phosphate availability may be a limiting factor for life in these worlds (e.g., [18,28,29]), cold subpolar regions could establish freeze-thaw cycles that concentrate such nutrients, possibly permitting microbial habitability [20]. Ocean worlds could be easily distinguished from classical HZ planets by their lower interior densities and from planets with rich H-He envelopes by their reduced atmospheric scale heights [20]. If such ocean worlds exist, they would prove that the carbonate-silicate cycle is not essential for habitability.

*What atmospheric compositions are best for characterizing the habitable zone?*
Even should the carbonate-silicate cycle be a relatively common biogeochemical cycle throughout the cosmos, are $CO_2$-$H_2O$ atmospheres necessarily common for HZ planets? For example, the idea that the classical HZ is for finding Earth-like life is not completely accurate. This is because $CO_2$ levels near the outer edge of the habitable zone are high enough to be toxic to Earth plants and animals (e.g. [2,30]). Thus, we should question whether the HZ outer edge should really be determined by the maximum greenhouse effect of $CO_2$.

This question is complex because planetary atmospheres contain additional greenhouse gases. Additional warming from secondary greenhouse gases (like $CH_4$ and $H_2$) would reduce the $CO_2$ levels required for surface habitability, possibly making them less toxic for some forms of life.

For instance, volcanoes, particularly on planets with reduced mantles, can outgas voluminous amounts of hydrogen to complement $CO_2$ and $H_2O$ warming, which can increase the HZ width by ~50% [12]. This idea had also been used to possibly explain the climate of early Mars (e.g. [31]). Another notion suggests that habitable planets can acquire large (tens to hundreds of bars) primordial $H_2$ envelopes, which would extend the outer edge of the solar system's HZ to nearly Saturn's orbit [11]. Such hydrogen-rich exoplanets (both $CO_2$-$H_2$ and primordial $H_2$ planets) are considerably easier to observe via transmission spectroscopy than classical $CO_2$ HZ planets [12,32], given the greater scale heights of the former.

Likewise, $CH_4$ is predicted to be a decent warming agent on HZ planets orbiting hotter stars (>~4500 K), but unlike other gases, it cools the surface for HZ planets orbiting cooler stars [13]. Recent 3-D simulations of the TRAPPIST-1 planets support the latter result [33]. Thus, HZ planets with high enough $CH_4$ levels orbiting cooler (mid-K to M) stars likely exhibit planetary surfaces that are too cold for life[13]. Such atmospheres would also be prone to atmospheric collapse [3].

Therefore, $CH_4$ may *not* be a good biosignature gas for HZ planets near the outer edge of cooler stars. Whereas methanogens are a good source of $CH_4$ on Earth, it is a mistake to suggest that this *must* hold for other planets (e.g., Titan's $CH_4$ is also abiotically-produced), especially around other stars. On the other hand, as long as $CH_4$ levels are not high enough to produce a cooling haze [34,35], relatively high (percent level) $CH_4$ concentrations on planets orbiting hotter stars *could* suggest inhabitance [13]. This is because such high $CH_4$ concentrations are very difficult to achieve without life [13,36], with the possible exception of ocean worlds [37].

Observations can reveal the effects of secondary greenhouse gases. Assuming that a universal carbonate-silicate cycle exists, discontinuities in the measured $pCO_2$ gradient from the classical inner to outer edge may be evidence of secondary greenhouse gases [3].

Although planets located near the outer edge of the $CO_2$-$H_2$ or $CO_2$-$CH_4$ HZ would be farther away and dimmer than those near the classical HZ outer edge (1.67 AU) (Figure 1)[2], the planet-star contrast ratio (defined as the ratio of planetary to stellar fluxes) at the $CO_2$-$H_2$ HZ outer edge (2.4 AU) declines by only a factor of $(2.4/1.67)^2$ ~2. Observing distances just beyond the outer limits of the classical HZ allows observations of TRAPPIST-1h, for instance [38]. It also permits sub-Earth sized planets near the classical HZ outer edge to be observed [3]. However, next-generation missions (and beyond) will need better wavefront stability to achieve such high contrast ratios. Ultimately, achieving this will require more investment in engineering and telescope design.

*Moist greenhouses and abiotic oxygen buildup*

At distances close to the inner edge, $H_2O$ molecules in the upper atmosphere dissociate rapidly enough to trigger a "moist greenhouse", desiccating a terrestrial planet with an Earth-like water inventory in ~ 4.5 Gyr (e.g., [2]). With a 1-bar $N_2$ (non-condensable) atmosphere on rapidly-rotating planets, this threshold occurs at a temperature of ~340 K (e.g. [2]). However, this threshold occurs at even lower temperatures at lower non-condensable inventories [39]. Such a process also leads to build up of abiotic oxygen, which has been argued to suggest uninhabitable conditions [39]. However, the mean surface temperatures of such planets can be clement ( e.g., < 300K for a 0.1 bar $N_2$ atmosphere), which suggests that life may be possible [3]. Missions should observe these planets to test both possibilities.

Some 3-D studies find that the moist greenhouse can be triggered at lower temperatures (< ~300 K) for tidally-locked planets ([40,41]). Given the uncertainties in the treatment of clouds, convection, heat transport, and radiative transfer in climate models, observations are needed to provide more information.

*The potential habitability of M-star systems*

M-stars are currently the most popular targets in the search for potentially habitable planets. This is in spite of a host of problems plaguing their planetary habitability including: 1) tidal-locking unless atmospheres are thick enough (e.g [42]), 2) intense space weather (e.g. [43,44]), 3) high impact velocities that erode planetary atmospheres [45,46], 4) a possibly high occurrence of controversial ocean worlds (e.g. [26,47], but also see ref: [48]) , 5) the cooling effect of $CH_4$ on such planets ([13] and see above), and 6) high pre-main-sequence superluminosity leading to early desiccation of HZ planets (e.g. [15,17]). This latter point is especially important for HZ definitions because most classical HZ studies have traditionally defined HZ boundaries for the zero-age or present-day HZ (e.g. [2,4]). However, many current M-star HZ planets were not in the HZ for the first tens of millions to up to *billions* of years after stellar formation, as they would have experienced a runaway greenhouse episode.

Thus, whereas the classical HZ suggests that M-star HZ planets could be habitable, the pre-main-sequence temporal evolution would actually predict such worlds to be uninhabitable deserts [15,17,48], unless they had acquired sufficiently large stores of water (tens to hundreds of Earth oceans or more) during accretion (e.g., [47]), migrated inward after the superluminous phase ended (e.g., [45]), or were replenished in a subsequent "late heavy bombardment"-type event [15]. Earlier works had also suggested that abiotic oxygen may build up to higher levels on HZ planets orbiting pre-main-sequence M-stars than any other star type (e.g. [17]) although magma oceans and atmospheric escape can remove such oxygen (e.g. [43,49–51]). For these reasons, we need to consider the temporal evolution of the HZ in assessments of habitability (e.g., [6,15]).

Nevertheless, M-stars should continue to be searched as they make up over half of our galaxy's stellar population (e.g., [52]). Their improved planet-star contrast also makes their planets easiest to detect via current methods (e.g. transit, direct imaging, and radial velocity). Thus, next-generation ground-based and space-based telescopes can produce abundant data on planets orbiting M-stars (e.g., [53]). Such missions would determine if M-star HZ planets have water vapor in their atmospheres, if any atmosphere at all.

*The case for observing A-star systems*

A-stars *may* harbor potentially habitable planets even though little has been published about their habitability. This is partially because they are far less common than M-stars, but also because the classical HZ limits had only been defined for F - M stellar systems (e.g., [2,4]), until recently [12,13]. However, there are reasons to *also* observe A-stars in spite of their fewer numbers, as next-generation missions plan to do [54]. One problem with A-star systems is their short main-sequence lifetime (<~ 2Gyr). However, microbial life on our planet took no more than ~700 Myr to arise [55], roughly equivalent to the main-sequence lifetime of an ~A3 star (~10,000 K)(e.g., [3]). These lifetimes are *still* conservative as habitable conditions on our planet may have already been in place less than 200 Myr after its formation [56,57], far shorter than A-star main-sequence lifetimes. With their short main-sequence lifetimes, HZ planets may also shed light on the origin of life.

Moreover, A-star HZ planets do not share some of the problems suffered by HZ planets orbiting M-stars. For instance, A-stars are *subluminous* during their pre-main-sequence, which favors water retention. Planetary orbits for these systems are also wider and more distant, which favors lower, more clement impact velocities. Likewise, close-in tidally-locked HZ planets should be very unlikely. If ocean worlds are particularly common around M-dwarfs (see ref:[47] for reasons), that would suggest that they are less common around A-stars. Also, methane has the strongest warming effect on these planets, resulting in a relatively wider $CO_2$-$CH_4$ HZ as compared to any other star type [13]. Plus, anti-greenhouse hazes may not form on planets orbiting A-stars (or F-stars) [58]. Although A-stars emit high levels of far-UV radiation, which suggests intense surface radiation environments ([59,60]), this does not preclude the existence of underwater or subsurface organisms. Their higher UV-C fluxes may also be favorable to prebiotic chemistry [61].

**CONCLUDING THOUGHTS**

To conclude, we have provided several predictions expected by HZ theory (i.e. the universality of the carbonate-silicate cycle, $CO_2$-$H_2O$ atmospheres, runaway greenhouses and abiotic $O_2$ buildup, and the potential habitability of A- and M-star systems) and the observations needed to test them. Only observations can improve how we use the HZ or discard it in favor of better search metrics. Testing these theoretical predictions would require greater investments in telescope design, which would lead to (for example) the increase in wavefront stability that is needed to observe smaller and more distant terrestrial planets.

For instance, a near-future architecture may find ~50 planets in the HZ of A - M stars (e.g., [54,62]), which are great numbers for some statistical analyses (e.g., [63]). With subsequent missions, we should aim to find hundreds, if not *thousands* of HZ planets to maximize the data return and improve our chances of finding life elsewhere. In time, the HZ can evolve from being a mere selection tool to becoming a true navigational filter capable of determining the HZ planets that are least likely to host life [3]. The gradual increase in data also improves our understanding of planetary atmospheres, which would be beneficial to all of astrophysics.

Finally, we should be careful about using our Earth to extrapolate about life on other planets, particularly those around other stars. The future of habitability studies will require first principles approaches where the temporal, spatial, geological, astronomical, atmospheric, and biological aspects of a planet's evolution are dynamically coupled [3,64]. This, together with improved observations, is the key to making more informed assessments. In turn, only through better observations can we improve such theoretical models [65].

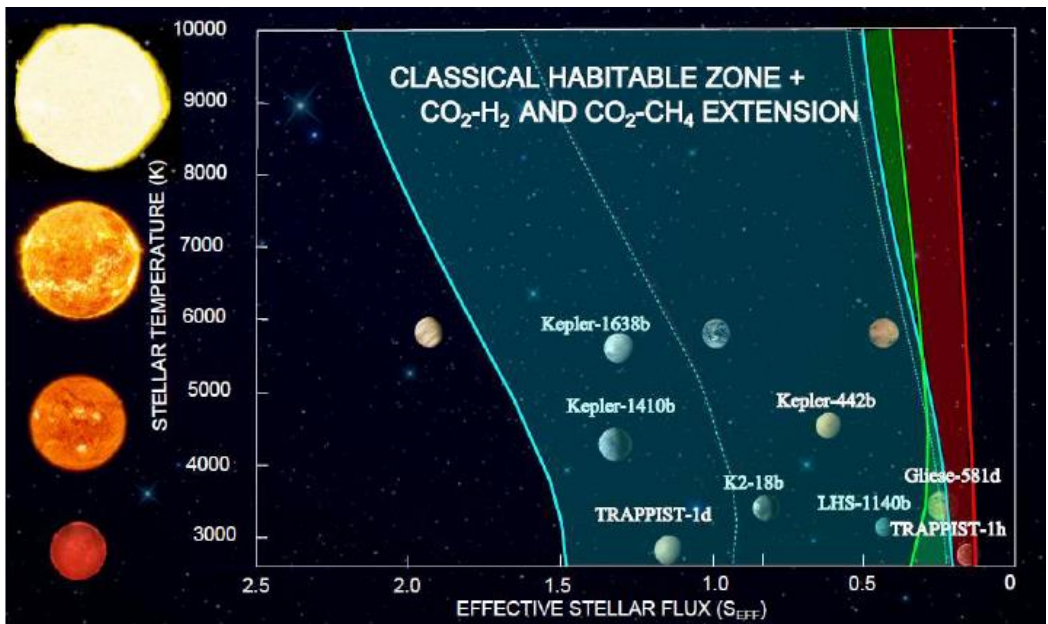

**Figure 1:** Various representations of the $CO_2$-$H_2O$ HZ. The classical $CO_2$ HZ boundaries (blue), extended for A- M stars (2,600 – 10,000 K)[13] shown alongside the extensions provided by the $CO_2$-$H_2$ (red)[12] and $CO_2$-$CH_4$ (green)[13] alternative HZ definitions. Some exoplanets and solar system planets are also shown. Reproduced from ref:[3].